\documentclass[longbibliography,showpacs,floatfix,nofootinbib,superscriptaddress,twocolumn,aps,prb]{revtex4-2}

\usepackage[T1]{fontenc}
\usepackage[utf8]{inputenc}
\usepackage{graphicx,color}
\usepackage{amsfonts}
\usepackage[figuresright]{rotating}  
\usepackage{amssymb}
\usepackage{amsmath}
\usepackage{mathtools}
\usepackage{psfrag}
\usepackage{tabularx}
\usepackage{textcomp}
\usepackage{units}
\usepackage{graphicx}
\usepackage{physics}
\usepackage{times}
\usepackage{ulem}
\usepackage[colorlinks=true,citecolor=blue,linkcolor=magenta,hypertexnames=false]{hyperref}
\usepackage{enumitem}
\usepackage[caption=false]{subfig}
\usepackage{tikz}
\usepackage{appendix}
\usepackage{dsfont}
\usepackage{bm}
\usepackage{bbm}
\usepackage{bbold}

\DeclareSymbolFont{bboldfont}{U}{bbold}{m}{n}
\DeclareMathSymbol{\bbzero}{\mathord}{bboldfont}{"30}

\newcommand{\Inp}[1]{\Big\langle #1\Big\rangle}

\newcommand{\LL}{\mathcal{L}}

\newcommand{\DD}{\mathcal{D}}

\newcommand{\FF}{\mathcal{F}}

\newcommand{\ii}{\text{i}}
\newcommand{\ee}{\text{e}}

\newcommand{\e}[1]{\text{e}^{#1}\,}

\begin{document}
\title{From weakly to strongly-interacting driven-dissipative bosons in one dimension}

\author{M. Z\"undel}
\affiliation{Universit\'e Grenoble Alpes, CNRS, LPMMC, 38000 Grenoble, France}
\affiliation{Universit\"at Innsbruck, Institut f\"ur Theoretische Physik, Technikerstra{\ss}e 21a, 6020 Innsbruck, Austria}
\author{L. Herviou}
\author{L. Canet}
\author{A. Minguzzi}
\affiliation{Universit\'e Grenoble Alpes, CNRS, LPMMC, 38000 Grenoble, France}

\begin{abstract}
	We consider a   one-dimensional driven-dissipative Bose-Hubbard model subjected to incoherent pump and one- and two-body losses, and analyze it by studying its two-point space-time correlations. By employing a combination of numerical methods, such as stochastic semi-classical simulations and  tensor network methods, as well as perturbation theory within the Keldysh formalism, we characterize the system at varying filling and interaction strength. We present results for 
	two complementary regimes: i) at large filling and weak interactions, where we show that Kardar-Parisi-Zhang scaling is visible in the linewidth of the spectral function and ii) at weak filling and strong interactions, where we analyze what remains of the mean field transition and identify a change of nature of the excitations. Our study covers two important regions of the phase diagram of such a system.
\end{abstract}

\maketitle

\section{Introduction}

Open bosonic quantum systems on a lattice model  a wide class of physical systems, ranging from coupled photonic Kerr cavities  to exciton polaritons in micropillars and Josephson-junction arrays. At weak interactions, these systems  have been the object of intense theoretical work, based on mean-field~\cite{CarusottoCiutiRMP}, truncated-Wigner methods~\cite{Carusotto2005PRB,Wouters2009PRB} and positive-P representation~\cite{Deuar2019PRXQ}. 
On the other hand, strongly interacting regimes are very promising for applications to quantum state manipulation and are needed for the realization of non-classical states of light, but notably challenging to describe theoretically. Several approaches have been put forward, including Monte-Carlo wavefunctions~\cite{Dalibard1992PRL,Molmer:93,Daley04032014},  matrix-product states (for the one-dimensional case)~\cite{Weimer2021RMP}, Gutzwiller and cluster mean field methods~\cite{Leboite2013PRL,Jin2016PRX}, diagrammatic expansions \cite{Li2016PRX,Biella2018PRB}, corner-space methods~\cite{Finazzi2015PRL}, and neural networks~\cite{HartmannCarleo2019PRL,NagySavona2019PRL,VicentiniBiella2019PRL,Yoshioka2019PRB}.
While several of the above methods have mainly dealt with local correlations of a small number of degrees of freedom (e.g. the single-sited Kerr resonator), the description of extended, open many-body systems remains an open challenge.

We focus here on the spectral function, related to the Fourier transform of the retarded Green's function, which provides information on the single-particle excitations arising in a system by adding or subtracting a particle to it. In many-body and open quantum systems, the spectral function generally displays broad features due to both interactions and drive/dissipation. In a previous study~\cite{Zuendel25SciPost}, we have obtained the two-point retarded Green's function for various interaction regimes of a Bose gas on a lattice in one dimension under Markovian one-body pump and losses. In that case, the steady state is an effectively infinite-temperature state fixed by the ratio of loss to pump rate, and corresponds to a dressed vacuum state.

In this work, we consider a different situation, obtained when the Bose gas is additionally subjected to two-body losses. Effective two-body losses can arise experimentally from pump saturation~\cite{CarusottoCiutiRMP}. The nonlinear losses saturate the incoherent pumping and stabilize a NESS even above the linear gain--loss threshold, where the model containing only one-body processes would be unstable. At mean-field level, this model exhibits a transition at $\gamma_p=\gamma_l$ between solutions with zero and non-zero density. At the single-site level, this combination of incoherent one-body pumping with one- and two-body losses realizes a quantum van der Pol oscillator~\cite{Dutta2019}. In zero dimension, the NESS is known exactly and has a finite occupation on both sides of the threshold, and a sharp dissipative phase transition emerges only in the large-occupation limit $\gamma_{2l}/\gamma_p\to0$~\cite{Li2025}. We comparatively explore both the weakly and strongly interacting regimes of the one-dimensional many-body problem.

In the weakly interacting semi-classical regime, the balance between non-resonant pumping and one- and two-body losses supports a stationary condensed state with spontaneously broken $U(1)$ symmetry. Its dynamics is accurately described by the stochastic driven-dissipative Gross--Pitaevskii equation. In one dimension, in an appropriate parameter regime, the long-wavelength phase fluctuations exhibit Kardar--Parisi--Zhang (KPZ) universal scaling~\cite{KPZ-original,Grindstein1993,Altman2015,sieberer_keldysh_2016}. The emergence of KPZ universality has been observed experimentally in a one-dimensional exciton-polariton condensate~\cite{Bloch2022}.
In this regime, the first-order correlation function follows a stretched-exponential decay in space and time~\cite{He2015}, evidenced as well in the temporal scaling of the correlations in momentum space~\cite{Wouters2015KPZ}. Here, by employing a semi-classical description, we show that  the spectral function allows to extract the dynamical critical exponent from the width of the dispersion line, which changes from the dispersive behavior $z=2$ predicted by the Bogoliubov theory to the KPZ prediction $z=3/2$.
 
In the strongly-interacting regime, we  provide a  fully quantum description of the  open Bose Hubbard model using a tensor network approach that we have specifically developed. First of all, in contrast to Ref.~\cite{Zuendel25SciPost}, for the case of two-body losses the non-equilibrium steady state (NESS) is not known. We compute it by representing the density with a matrix-product-operator (MPO) and using density matrix renormalization group (DMRG)~\cite{White1992,White1993,Ostlund1995,Verstraete2004,Schollwock2005, Schollwock2011} as done in Ref.~\cite{MascarenhasFlayacSavona2015} to minimize the Liouvillian spectral gap.
Vectorizing the obtained density matrix, we describe the time-evolution of excited states with time-evolving block-decimation (TEBD)~\cite{Vidal2004}, time-dependent variational principle (TDVP)~\cite{Haegeman2011} and an exponential approximation to the time-evolution~\cite{Zaletel15, VanDamme2024}. This allows us to reach long simulation times with controllable error and to obtain the retarded Green's function and the spectral function.
We show results for the spectral function from the fully quantum model, both at weak and at strong interactions, highlighting the differences with respect to the large-filling semi-classical regime. We show that the decay of the retarded Green's function at weak interaction is well captured by a one-loop calculation performed within the Schwinger-Keldysh formalism.
We then analyze the lowest lying excitation in the Liouvillian spectrum. While there is no dissipative phase transition in the quantum model when changing the parameters across the mean-field transition point, we report a transition in the  first excited state, which changes from localized to extended, as demonstrated by analyzing the form factors of this excitation.

The paper is organized as follows.
Starting from the Liouvillian evolution of the Bose-Hubbard model (BHM) in second quantization, we derive in Section~\ref{sec_methods} the Keldysh action and the semi-classical driven-dissipative Gross-Pitaevskii equation (GPE) and introduce the relevant observables.
In Section~\ref{sec_semicl}, we perform the mapping to the semi-classical evolution and characterize the spectral function in this regime.
In Section~\ref{sec_quantumdyn}, we analyze the open BHM by means of tensor networks. We describe our algorithm and provide results for the spectral function both at strong and at weak interactions.
Finally Section~\ref{sec_conclusions} contains our concluding remarks and outlook.

\section{Model and definitions}
\label{sec_methods}
We consider a periodic lattice of $L$ sites,  with  lattice spacing $a$, occupied by  bosons. Their unitary time evolution  is described by  the Bose Hubbard Hamiltonian 
\begin{align}\label{BH_hamiltonian}
	H = \sum_{i=1}^{L} \Big[-J\big(b_i^{\dagger}b_{i+1}+\text{h.c.}\big) + \frac{U}{2}b_i^{\dagger}b_i^{\dagger} b_i b_i\Big]\;,
\end{align}
which consists of a quadratic term proportional to the hopping energy $J$ and an on-site two-body interaction of strength $U$. We use  periodic boundary conditions $b_{i+L}\equiv b_i$.
The bosonic creation and annihilation operators fulfill the usual commutation relations $[b_i, b_j^{\dagger}]= \delta_{ij}$, $[b_i, b_j] = 0 = [b_i^{\dagger}, b_j^{\dagger}]$.
Each site is homogeneously coupled to Markovian baths describing the incoherent gain and loss of particles.  The full quantum evolution of the density matrix is described by the Gorini–Kossakowski–Sudarshan master equation: a Lindblad master equation 
$\partial_t \rho=\LL \rho$, with 
\begin{align}
\label{eq_lindblad}
\LL\rho &= -\ii [H, \rho] + \gamma_l\sum_{i = 1}^L \DD [b_i]\rho + \gamma_p\sum_{i = 1}^L \DD [b_i^{\dagger}]\rho \nonumber\\
&\quad
+ \gamma_{2l} \sum_{i=1}^L \DD[b_i^2]\rho
+\gamma_d \sum_{i=1}^L\DD[b_i - b_{i+1}]\rho
\;,
\end{align}
where the dissipators are defined as
\begin{align}
\DD[L_i]\rho = L_i \rho L_i^{\dagger} - \frac{1}{2} \{L_i^{\dagger} L_i, \rho\} \;,
\end{align}
with $L_i=b_i$ for the homogeneous, on-site one-particle loss, $L_i=b_i^\dagger$  for the  one-particle pump, and $L_i=b_i^2$ for the  two-particle loss, with respective rates $\gamma_l$, $\gamma_p$, and $\gamma_{2l}$. Additionally, we introduced a one-particle loss term with support on two neighboring sites,  $L_i = b_i - b_{i+1}$ at a rate $\gamma_d$, which encodes a dissipative hopping, modeling a momentum-dependent loss rate.

\subsection{Keldysh action}
We follow Ref.~\cite{sieberer_keldysh_2016} to derive from the Lindblad master equation~\eqref{eq_lindblad} the microscopic action in the Keldysh rotated basis, in terms of classical and quantum fields $\varphi_{c/q,j}=\frac{1}{\sqrt{2}}(\varphi_{+,j}\pm\varphi_{-,j})$ on each lattice site $j$, given by
\onecolumngrid
\begin{align}\label{action}
	S = \int \dd t &\sum_j \Bigg\{
	\varphi_{c,j}^* \, \ii (\partial_t - \kappa_0) \varphi_{q,j} + \varphi_{q,j}^* \, \ii (\partial_t + \kappa_0) \varphi_{c,j} + \ii \gamma_0 |\varphi_{q,j}|^2\\
	&-\frac{\ii\gamma_d}{2}
    \Big(
        (\Delta\varphi_{q,j})^*
        \Delta\varphi_{c,j}
        -
        (\Delta\varphi_{c,j})^*
        \Delta\varphi_{q,j}
        +
        2\lvert\Delta\varphi_{q,j}\rvert^2
    \Big)\nonumber\\
	&+ J \Big(
	\varphi_{c,j}^*\varphi_{q,j-1} + \varphi_{c,j}^* \varphi_{q,j+1} + \varphi_{q,j}^* \varphi_{c,j-1}  + \varphi_{q,j}^* \varphi_{c,j+1}
	\Big)\nonumber\\
	&- \frac{U}{2} \Big(
	(\varphi_{c,j}^2 + \varphi_{q,j}^2) \varphi_{c,j}^* \varphi_{q,j}^* + (\varphi_{c,j}^{*2} + \varphi_{q,j}^{*2}) \varphi_{c,j} \varphi_{q,j}
	\Big)\nonumber\\
    &- \ii \frac{\gamma_{2l}}{2} \Big(
	- (\varphi_{c,j}^2 + \varphi_{q,j}^2) \varphi_{c,j}^* \varphi_{q,j}^* + (\varphi_{c,j}^{*2} + \varphi_{q,j}^{*2}) \varphi_{c,j} \varphi_{q,j} - 4 \varphi_{c,j} \varphi_{q,j} \varphi_{c,j}^* \varphi_{q,j}^*\Big)
	\Bigg\}\nonumber\;,
\end{align}
where $\Delta\varphi_{\alpha,j}\equiv\varphi_{\alpha,j}-\varphi_{\alpha,j+1}$ with $\alpha=c,q$, and  $\kappa_0 \equiv (\gamma_l-\gamma_p)/2$ and $\gamma_0\equiv \gamma_l+\gamma_p$.
\newpage

\twocolumngrid
\noindent

\subsection{Observables}
In the following, we are interested in the spectral function
\begin{align}\label{spectral}
  A(\omega, k) = - \frac{1}{\pi} {\rm Im} \, G^R(\omega, k) \;,
\end{align}
which is related to the Fourier transform of the retarded Green's function, defined on the lattice by
\begin{align}
  \label{G_ret}
  G^R_{j\ell}(t,t') = -\ii \Theta(t-t') \Inp{[b_j(t), b_\ell^{\dagger}(t')]}\;,
\end{align}
where $\langle ... \rangle $ indicates the trace weighted with the NESS 
\begin{align}
\Inp{b_j^{\dagger}(t)b_0}= {\rm Tr} \Big\{ b_j^{\dagger} \e{\LL t} \Big[ b_0 \rho_{\text{NESS}}\Big]
\Big\}\;.
\end{align}

This quantity is closely related to what is observed in quantum  optics experiments, where one measures the first-order correlation function
\begin{align}\label{eq_g1_quantum}
    g^{(1)}(x, t; x',t')=\frac{\left\langle
    \Psi^\dagger\left(x, t\right) \Psi\left(x', t' \right)
    \right\rangle}{\sqrt{\left\langle\rho\left(x, t \right)\right\rangle} \sqrt{\left\langle\rho\left(x', t'\right)\right\rangle}}\;.
\end{align}
expressed in terms of the  bosonic field operators $\Psi(x)$, $\Psi^\dagger(x)$, here expressed for continuous space.

In the semi-classical limit, which is introduced in the next section, the corresponding first-order correlation is expressed as
\begin{align}\label{eq_g1}
    g^{(1)}(x, t; x',t')=\frac{\left\langle
    \varphi^*\left(x, t\right) \varphi\left(x', t' \right)
    \right\rangle_\xi}{\sqrt{\left\langle\rho\left(x,  t\right)\right\rangle_\xi} \sqrt{\left\langle\rho\left(x', t'\right)\right\rangle_\xi}}\;.
\end{align}
where $\langle...\rangle_\xi$ indicates the average over the noise realizations,  $ \varphi(x, t)$ is the classical field, described by the driven-dissipative Gross-Pitaevskii equation (see Eq.(\ref{ddGPE}) below) and $\rho(x)=|\varphi(x)|^2$ is the density.
All correlation functions are evaluated within the stationary regime. Therefore we start their calculation  starting from the reference time $t_{\text{stat}}$ and  we abbreviate $g^{(1)}(x,t)=g^{(1)}(x,t_{\text{stat}}+t; 0, t_{\text{stat}})$ in the semi-classical limit.

In the quantum regime and in the lattice case, taking as reference point $(x',t')=(0,0)$, the numerator of (\ref{eq_g1_quantum}) requires as input the discrete two-point correlator
\begin{align}
    \rho^{(1)}_j(t) = \Inp{b_j^{\dagger}(t) b_0(0)} = \ii (G_{0j}^<(0,t))^*\;.
\end{align}
equivalent to the complex conjugate of the lesser Green's function $G^<$. In the  single-band description of the lattice, this  allows one to reconstruct the full 
correlations Eq.~(\ref{eq_g1_quantum}), expressed for the lattice case.

\section{Semi-classical regime}\label{sec_semicl}
At sufficiently large densities and weak interactions, driven-dissipative bosons can form a quasi condensate in one dimension, with large occupancy of a single mode, described by a classical field  $\varphi=\sqrt{n}e^{i \theta}$.
Its equation of motion is derived from a Hubbard-Stratonovich transformation of the semi-classical limit of the Keldysh action~\eqref{action} for $\varphi_c$, where all terms non-linear in the quantum field, except for the one purely quadratic in $\varphi_q$, are neglected. This corresponds to the Martin–Siggia–Rose–Janssen–De Dominicis formalism~\cite{MartinSiggiaRose1973,Janssen1976,DeDominicis1976}.

As the quadratic part in the quantum field $\gamma(k) = \gamma_0 - 2 \gamma_d [\cos(a k) - 1]$ is gapped since $\gamma_0=\gamma_p+\gamma_l\neq0$, we can safely ignore the momentum dependence of the noise strength in the following, setting $\gamma(k)=\gamma_0$. This is in contrast to Ref.~\cite{MarinoDiehl2016PRL,MarinoDiehl2016PRB}.

Recalling that the relation of the classical field to the coherent field is given by $\varphi_c\rightarrow\sqrt{2}\varphi$, we arrive at the driven-dissipative Gross-Pitaevskii equation, which on the lattice reads
\begin{align}\label{ddGPE}
    \ii \partial_t \varphi_{j}(t)
	&= \FF^{-1}\Big[E(k)- \ii\kappa(k)\Big] \varphi_{j}(t)\\
	&\quad+ (U - \ii \gamma_{2l}) |\varphi_j(t)|^2 \varphi_{j}(t) +\ii \xi_j(t)\nonumber\;,
\end{align}
with $\xi_j$ complex noises  Gaussian distributed, of zero mean and covariance
\begin{align}\label{noiseddGPE}
    \Inp{\xi^*_j(t) \xi_{j'}(t')} = 2 \sigma \delta_{j,j'} \delta(t-t')\;,
\end{align}
where $\sigma = \frac{\gamma_0}{4a}$, and 
\begin{align}
    E(k) &= - 2J \cos(a k)\;,\\
    \kappa(k) &= \kappa_0 - \gamma_d [\cos(a k)-1]\;.
\end{align}

\subsection{Emergence of KPZ universality from the  driven-dissipative Gross-Pitaevskii equation}
In certain conditions, the effective dynamics of the  phase $\theta(x,t)$ of the condensate has been shown~\cite{Grindstein1993,Altman2015,sieberer_keldysh_2016} to follow the KPZ equation, which reads 
\begin{align}
    \partial_t \theta = \nu\nabla^2\theta + \frac{\lambda}{2} (\nabla \theta)^2 + \eta\;,
\end{align}
with diffusion constant $\nu$, non-linearity $\lambda$ and $\eta$ a real Gaussian noise of zero mean and covariance
\begin{align}
    \langle \eta(x,t)\eta(x',t')\rangle = 2D\delta(x-x')\delta(t-t')\;,
\end{align}
with $D$ the noise strength. The three parameters of the effective KPZ equation are given in terms of the microscopic parameters of the driven-dissipative  Gross-Pitaevskii equation \eqref{ddGPE}-\eqref{noiseddGPE}~\cite{helluin2025phase}.
In this regime, the condensate phase correlation is expected to scale with universal power-law exponents according to 
\begin{align}
    C_{\theta\theta}^{\text{(KPZ)}}(x,t) 
    &= \langle[\theta(x,t)-\theta(0,0)]^2\rangle\\
    &\sim \begin{cases}
    &t^{2\beta}\,,\qquad x\rightarrow 0\;,\\
    &|x|^{2\chi}\,,\qquad t\rightarrow 0\;.
    \end{cases}
\end{align}
where in one dimension, the exponents are known exactly due to symmetries and are given by $\beta=\frac{1}{3}$, $\chi=\frac{1}{2}$, $z=\frac{\chi}{\beta}=\frac{3}{2}$.
Finally, for small density fluctuations,
the two-point function $g^{(1)}$ is entirely determined by the phase fluctuations,i.e.
\begin{align}
    |g^{(1)}(x,t)|\approx \e{-\frac{1}{2}C_{\theta\theta}(x,t)}\;,
\end{align} 
which means that the scaling of the phase-phase correlation function $C_{\theta \theta}$ can be read off from the logarithm of $|g^{(1)}|$.

\subsection{Scaling in the first-order correlation function}
In this section, we study the semi-classical dynamics, given by Eq.~\eqref{ddGPE} 
for a set of parameters such that the condensate is in the KPZ regime. The choice of parameters is important, since other regimes dominated by space-time vortices  and solitons are also found from Eq.~\eqref{ddGPE} \cite{Vercesi2023}.
Inspired by the experimental parameters~\cite{Bloch2022}, translated to the adiabatic regime, we choose a  low enough
value of the interaction strength to decrease the number of vortices and set in our simulations the parameters as 
\begin{align}\label{param}
    \gamma_l &= 0.07\,,\;\;\gamma_p = 0.0774 \,,\;\; \gamma_{2l} = 2.44 \cdot10^{-4}\,,\;\; \gamma_d = 1.27\,,\nonumber\\
    J &= 0.9\;,\quad U = 10^{-3} \;, \quad a=0.224 \;, \quad \Delta t = 0.02\;.
\end{align}
All quantities are dimensionless, with lengths expressed in units of $\ell=4.4\mu\textrm{m}$ and times in units of $\tau=1\textrm{ps}$.
The above choice for the non-linearity $U$ avoids the soliton regime that makes the steady-state condensate unstable to linear order and displays scaling properties different from KPZ~\cite{Vercesi2023}. 
Further, we note that compared to earlier theoretical works, we choose here microscopically realistic parameters.
In particular, we do not rescale the field as in Refs.~\cite{He2015, Squizzato2018}, or rescale the amplitude of the noise as in Ref.~\cite{Vercesi2023}. These parameters set the mean-field condensate density to $\rho_0\approx 15.2\,a^{-1}$.

\begin{figure}[htb]
\centering\hspace{-0.9em}
\includegraphics[width=1.03\linewidth]{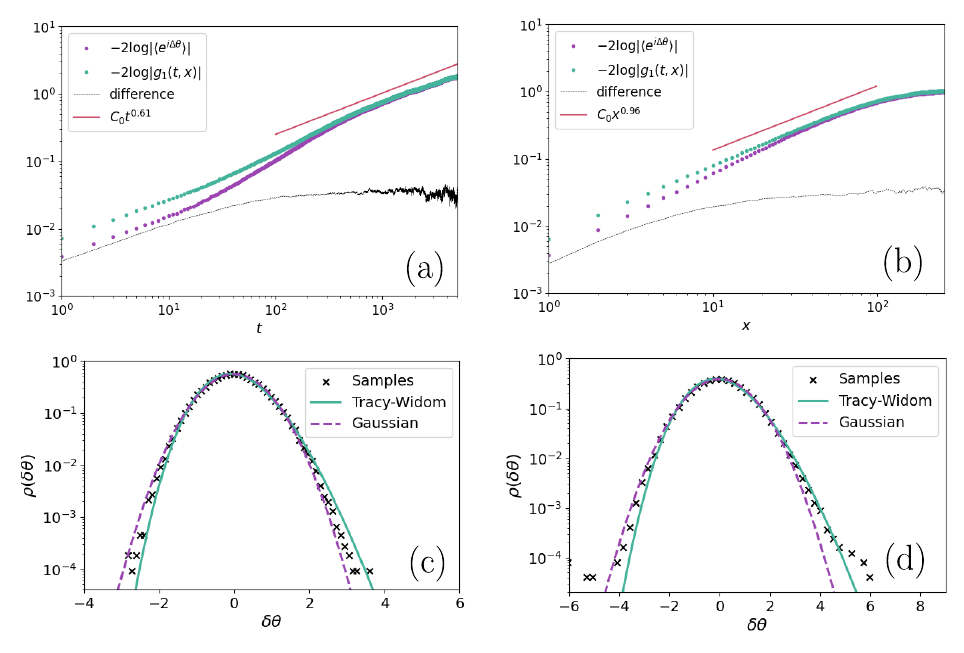}
\caption{(a) Temporal correlations $- 2 {\rm log} |g^{(1)}(0,t)|$ (turquoise)
evaluated with respect to the reference time $t_{\rm stat}=1000$ ps, as defined below Eq.~\ref{eq_g1}, where the system has already reached the non-equilibrium stationary state,  phase-phase correlator (purple) and their absolute difference (grey). (b) Spatial correlation $- 2 {\rm log} |g^{(1)}(x,0)|$, phase-phase correlations, and their absolute difference, with the same color code. (c)-(d): Statistics of the phase fluctuations for $n_{\text{traj}}=10^5$ trajectories; (c) at early times $t\approx 30$, statistics intermediate, and (d) at $t\approx 2000$ in the KPZ regime. The system size is taken $L=2^{10}$, the other parameters are given in Eq.~\eqref{param}. }\label{fig_kpz_scaling}
\end{figure}

We simulate the stochastic time evolution of the driven-dissipative Gross-Pitaevskii equation~\eqref{ddGPE} by means of a split-step algorithm, adding noise according to the Euler-Maruyama scheme~\cite{Maruyama1955} and in the end average over $n_{\rm traj}$ noise realizations. Starting from a finite density with spatially constant angular values, we wait for the system to reach its steady state (here, after $t_{\rm stat}=1000$ ps) and then calculate the normalized two-point correlation function $g^{(1)}(x,t;x't')$ defined in Eq.~\eqref{eq_g1}.

Fig.~\ref{fig_kpz_scaling} shows our results for the first-order space and time correlations as well as the probability distribution of the phase fluctuations for the parameters given in Eq.~\eqref{param}. We base our identification of the KPZ regime for the phase $\theta$ of the condensate on the following three observations:
\begin{enumerate}
    \item The correlation function $g^{(1)}$ is dominated by the phase fluctuations. Hence, phase-density and density-density correlators are negligible \cite{Deligiannis-EPL}. These are in fact the assumptions underlying the mapping of the phase dynamics to the KPZ equation.
    \item The logarithm of the $g^{(1)}$ correlation function scales in space and time with the expected critical  exponents of the KPZ universality class,   $\beta=1/3$ and $\chi=1/2$ ~\cite{He2015}.
    \item The statistics of the phase fluctuations are well described by the Tracy-Widom distribution ~\cite{Squizzato2018}.
\end{enumerate}
We observe on Fig.~\ref{fig_kpz_scaling}, that, starting from intermediate times $t\geq200$ ps, the behavior of the temporal $g^{(1)}$ correlation function is dominated by the phase fluctuations and displays a stretched exponential decay with $\beta\approx0.31$, which matches to a good accuracy with the KPZ scaling exponent $\beta=1/3$. The spatial $g^{(1)}$ correlation function also displays at large distances a stretched exponential decay with exponent $\chi=0.48$, which  is also close to the expected KPZ exponent $\chi=0.5$. Finally, we see that at large times ($t\approx 2000$ps) the statistics of the phase fluctuation at a single spatial point follows the Tracy-Widom (TW) distribution, best seen on the right tail, compared to the earlier times ($t\approx30$ ps) that display a more symmetric distribution,  in between Gaussian and TW statistics. This is to be expected, as at early times universal KPZ fluctuations have not fully  developed yet,  and  the density fluctuations still play an important role in this regime, as can be seen in Fig.~\ref{fig_kpz_scaling}~(a).

\begin{figure}[h]
\centering
\includegraphics[width=1.02\linewidth]{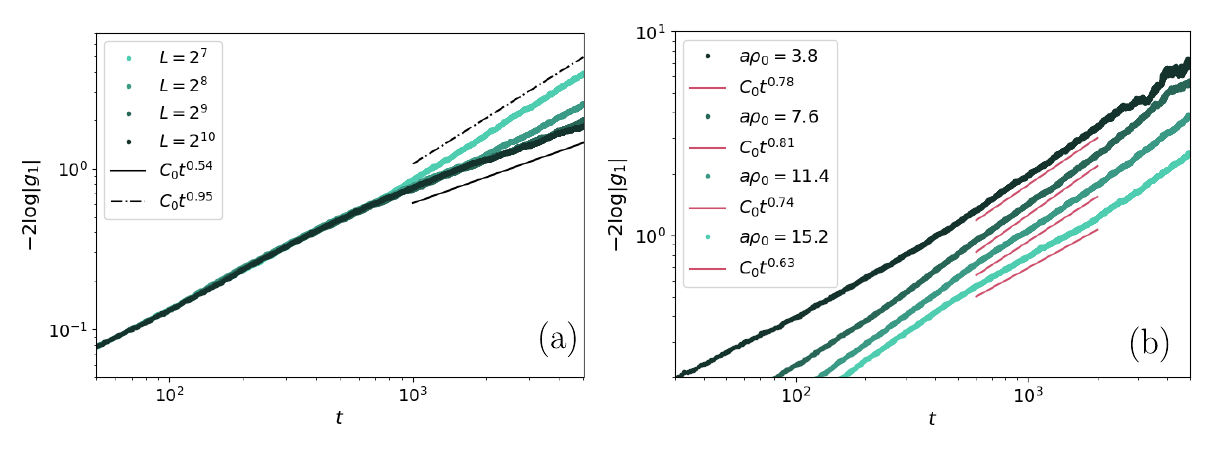}
\caption{The fading of the KPZ scaling towards small system sizes and low densities, keeping the remaining parameters as in Fig.~\ref{fig_kpz_scaling}. (a) Influence of the system size: temporal correlation for different system lengths $L=2^7, 2^8, 2^9, 2^{10}$, zoomed into the relevant time window. The KPZ scaling is still visible for $L=2^8$, but fades away for smaller system size $L=2^7$. (b) Influence of the mean density: Temporal correlation lowering the density by tuning the loss rate $\gamma_l=0.07,0.07185,0.0734,0.07525$ towards the pump threshold condition $\gamma_l=\gamma_p$. The system size is $L=2^8$.}\label{fig_kpz_lower}
\end{figure}

Starting from these parameters, we assess the robustness of the KPZ scaling towards the quantum regime, i.e. towards lower lattice occupation and smaller system sizes. In Fig.~\ref{fig_kpz_lower}~(a), one can observe that, when lowering the system size, the temporal scaling is no longer detectable for a system size of $L=2^7$, whereas for $L=2^8$ it can still be seen. For this system size, we lower the density by tuning the loss rate $\gamma_l$ towards the value of the pump $\gamma_p$, which is displayed in Fig.~\ref{fig_kpz_lower}~(b). We find that already for $25{\%}$ less particles, space-time vortices occur and the KPZ scaling is lost towards lower densities.
As discussed in Appendix~\ref{sec_qm_to_cl}, accessing  the semiclassical KPZ regime from the simulations of the microscopic quantum dynamics is hindered not only by the required system sizes and evolution times, but also by a qualitative difference between the quantum and semiclassical steady states. In our simulations of the quantum system, the filling remains low and largely insensitive to the distance from the mean-field threshold, indicating that if a similar crossover exists as in the semi-classical simulations then it is probably at different parameters.

\onecolumngrid
\subsection{Spectral function in the semi-classical regime}
\label{kpz-width}
Before presenting the results of the semi-classical simulations for the spectral function, let us briefly  recall the results of the excitation spectrum obtained within the Bogoliubov  theory. The latter is readily  derived from the Keldysh action by considering the bare inverse Green's function evaluated at the mean-field level in the broken-symmetry phase, corresponding to  non-zero bosonic density $\rho_0$.
We write the classical field in the frame rotating with $\mu$ as
\begin{align}
    \varphi_{c,j}(t)
    &=
    \ee^{-\ii\mu t}
    \left[
        \sqrt{2}\phi_0
        +
        \delta\varphi_{c,j}(t)
    \right]\;.
\end{align}
The homogeneous saddle is time independent and given by $\varphi_c=\sqrt{2}\phi_0$ and $\varphi_q=0$, with $|\phi_0|^2\equiv\rho_0=-\kappa_0/\gamma_{2l}=(\mu+2J)/U$.
Ordering the fields as $\varphi=(\varphi_c,\varphi_c^*,\varphi_q,\varphi_q^*)^T$, we define the different components of the bare inverse Green's function as 
\begin{align}
	\frac{\delta^2 S[\varphi]}{\delta \varphi^*_{a,i}(t) \delta \varphi_{b,j}(t')} \,  \Big|_{\varphi_{c}=\sqrt{2}\phi_0,\varphi_q=0}
	\equiv
	\begin{pmatrix}
		{\bbzero}_{2\times2} && {\mathbb P}^A_{i,j}(t,t')\\
		{\mathbb P}^R_{i,j}(t,t') && {\mathbb P}^K_{i,j}(t,t')
	\end{pmatrix}\;.
\end{align}
The corresponding inverse retarded Green's function $\mathbb P^R_{i,j}(t,t')$ in the Bogoliubov approximation  is given in Fourier space by:
\begin{align}
	\mathbb P^R(\omega, k) =
	\begin{pmatrix}
		\omega - E(k)  + \mu + \ii\kappa(k) - 2(U-\ii\gamma_{2l}) \rho_0 && - (U-\ii\gamma_{2l}) \phi_0^2\\
		- (U+\ii\gamma_{2l}) \phi_0^{*2} && -\omega - E(k) + \mu -\ii\kappa(k) - 2(U+\ii\gamma_{2l}) \rho_0
	\end{pmatrix}\,.
\end{align}
The excitation spectrum is obtained  from the solutions of  $\det \mathbb P^R(\Omega(k),k)=0$, yielding
\begin{align}\label{eq_omega_bog}
	\Omega_{\pm}(k) = \ii(\kappa_0+ \gamma_d(\cos(ka)-1))\pm\sqrt{2J(\cos(ka)-1)\Big[2J(\cos(ka)-1)-2U\rho_0\Big]-\gamma_{2l}^2\rho_0^2}\;.
\end{align}
\twocolumngrid

\begin{figure}[h]
\centering
\includegraphics[width=.9\linewidth]{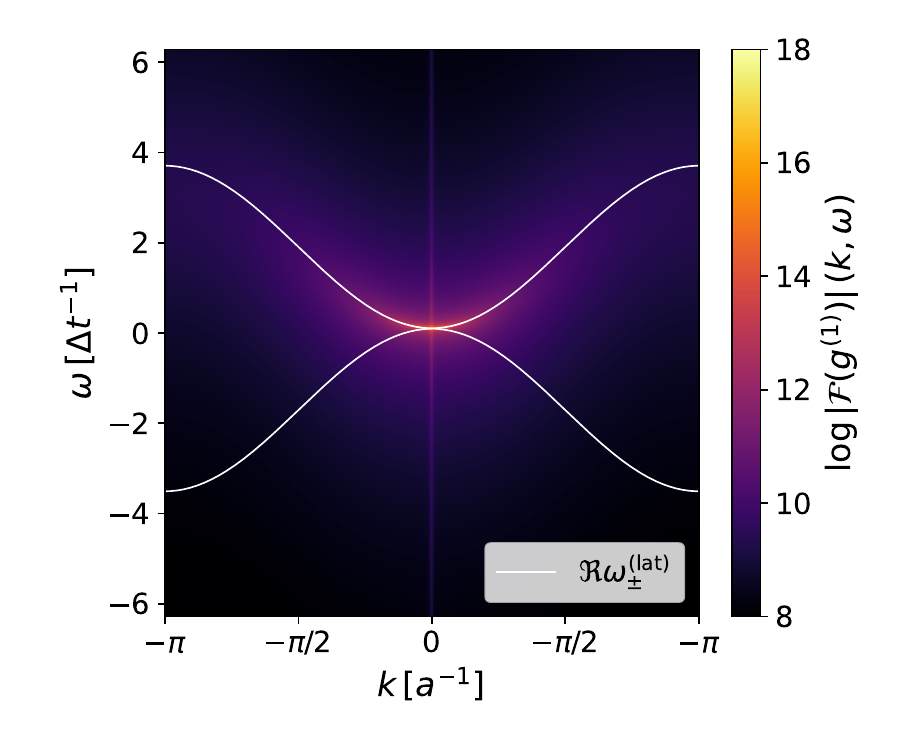}
\caption{Spectral function from semi-classical simulation, in the frequency-momentum plane, with wavevector in units of inverse lattice spacing $a^{-1}$ and frequency in units of inverse time step $\Delta t$. The  parameters are given in~\eqref{param} and system length $L=2^{10}$. The result is overlaid with the Bogoliubov prediction~\eqref{eq_omega_bog} for the excitation dispersion of lattice bosons (white line).}
\label{fig_spectrum_semicl}
\end{figure}

\begin{figure}[h]
\centering
\includegraphics[width=.47\linewidth]{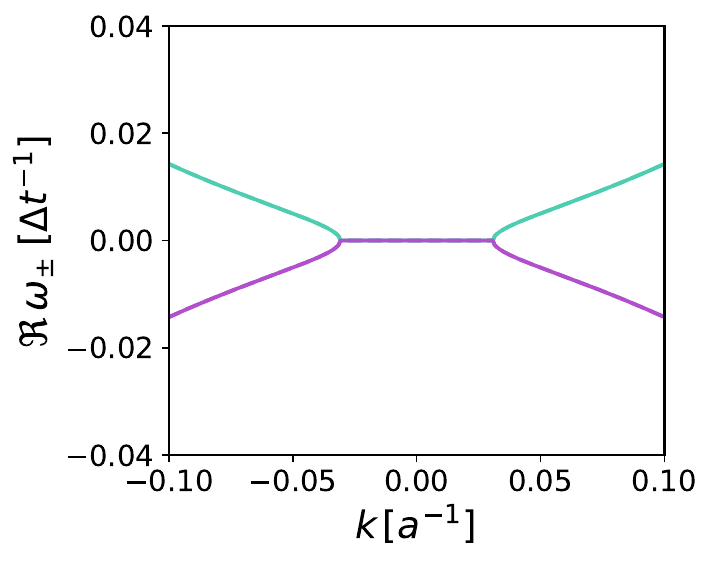}
\includegraphics[width=.47\linewidth]{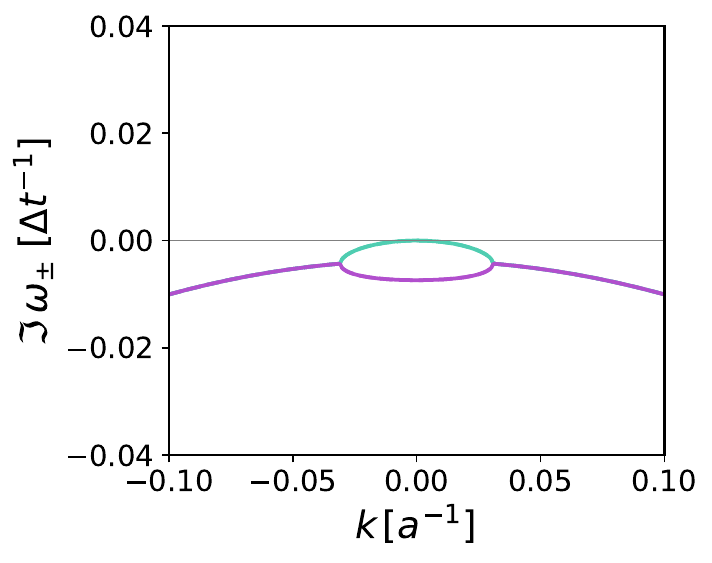}
\caption{Zoom on real (left panel) and imaginary (right panel) part of the Bogoliubov excitation dispersion $\Omega_\pm(k)$ given in Eq.~\eqref{eq_omega_bog}, near $k\approx0$, with frequency in units of the inverse time step $\Delta t^{-1}$ and wavevector in units of inverse lattice spacing $a^{-1}$.
In the right panel, the linewidth of one of the two excitation branches vanishes at $k=0$, corresponding to the dissipative Goldstone mode associated with the broken $U(1)$ symmetry in the out-of-equilibrium theory.}\label{fig_Bog}
\end{figure}

In Fig.~\ref{fig_spectrum_semicl}, the spectrum obtained from the semi-classical simulation is displayed, overlaid with the Bogoliubov prediction~\eqref{eq_omega_bog}, which  describes well  the curvature of the dispersion. Further, we show in Fig.~\ref{fig_Bog} the behavior near $k=0$ of the positive and negative frequency branches. The real part gives the excitation dispersion, also shown in Fig.~\ref{fig_spectrum_semicl}, the imaginary part shows the mean field prediction of the width of the excitation branches. Making use of the mean-field solution $\rho_0=-\frac{\kappa_0}{\gamma_{2l}}$, one readily checks that the imaginary part of one of the excitation branches vanishes at $k=0$.  
The closure of the dissipative gap of one of the branches indicates the broken $U(1)$ symmetry and the associated dissipative Goldstone modes~\cite{wouters_excitations_2007,Claude2025}.

However, the Bogoliubov theory, that describes only linear fluctuations around the mean field solution, fails to capture the growth of phase fluctuations at long times and the emergence of the KPZ universal scalings.
Under the assumption
\begin{align}
    \e{-\frac{1}{2}C_{\theta\theta}(x,t)} \approx 1 -\frac{1}{2}C_{\theta\theta}(x,t)\;,
\end{align}
we can relate the width of the spectral function peak to the scaling of the two-point phase-phase correlation function, implying that we expect  $A(k, \omega)\propto \ii k^z$,  with dynamical critical exponent $z=3/2$ in the KPZ regime. In Fig.~\ref{fig_z_from_spectrum}, we compare the fitted width of the dispersion branches from the semi-classical evolution with the Bogoliubov prediction. The Bogoliubov branches are asymptotically free and match the quadratic increase of the linewidth with momenta (dashed line), whereas the fitted width of the semi-classical spectral function follows more closely the KPZ predictions (solid line). We note that if the correspondence to the KPZ scaling function would be exact, then it would follow the standard scaling form~\cite{Praehofer2004}, where for a dispersive fluid one has to consider the fluctuations on top of the dispersion branch $\omega(k)$~\cite{Kulkarni2013}
\begin{equation}
    C_{\theta\theta}(k,\omega)= |k|^{-(2\chi+d+z)}\mathcal F\Big(\frac{(\omega-\omega(k))}{|k|^z}\Big)\;,
    \label{eq:KPZ-collapse}
\end{equation}
where the scaling of the amplitude gives $2\chi+d+z=7/2$ in one dimension.

We display in Fig.~\ref{fig_z_from_spectrum} the collapse of the spectra at various wavevectors according to Eq.~\eqref{eq:KPZ-collapse}. We obtain a good agreement with the KPZ scaling in the variable $(\omega-\omega_k)/|k|^z$ for a broad range of $k$ values\footnote{We also notice that the scaling amplitude prefactor $k^\alpha$ does not match the predicted pure-KPZ behavior. We attribute this to the fact that the KPZ regime only emerges over a finite window of momenta and frequencies, and it does not extend in particular to the small-$k$ behavior.}. Our results confirm the spectral function as additional, experimentally accessible probe of the KPZ regime in exciton-polariton condensates.  

\begin{figure}[h]
\centering
\includegraphics[width=.9\linewidth]{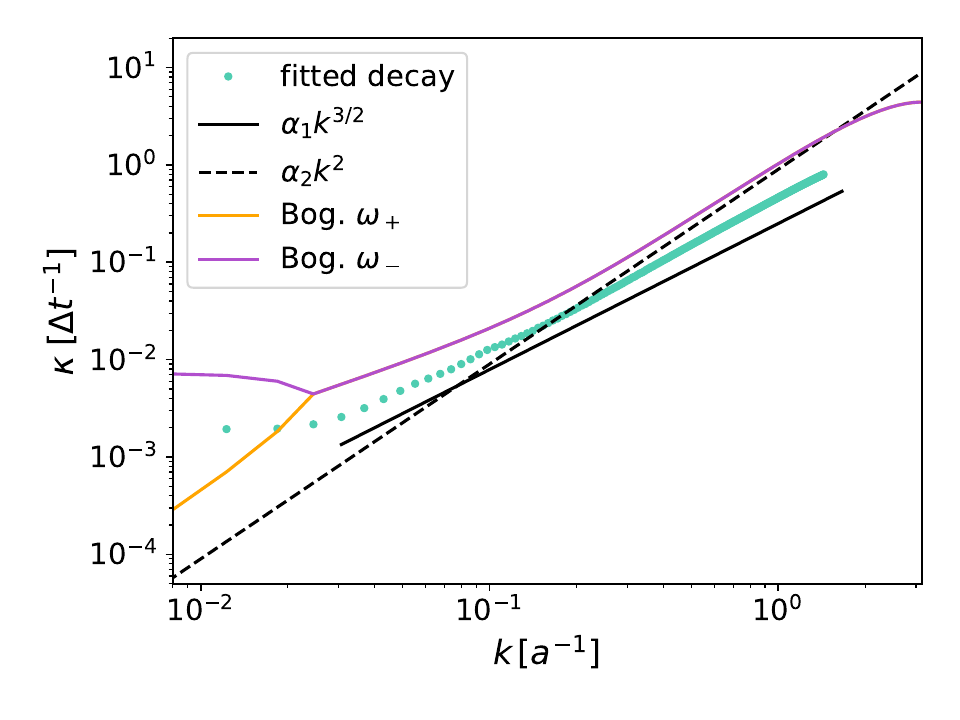}
\caption{Width of spectral lines extracted from the spectrum shown in Fig.~\ref{fig_spectrum_semicl}
for the parameters in~\eqref{param} (dots) and from the Bogoliubov theory (orange and magenta solid lines). The width are obtained by fitting Lorentzian curves for each $k$-slice. The black dashed line indicates the $\kappa(k)\sim k^2$ behavior predicted by Bogoliubov theory and the black solid line indicates the $\kappa(k)\sim k^{3/2}$ behavior expected for the KPZ regime.}
\label{fig_z_from_spectrum}
\end{figure}

\begin{figure}[h]
\centering
\includegraphics[width=.8\linewidth]{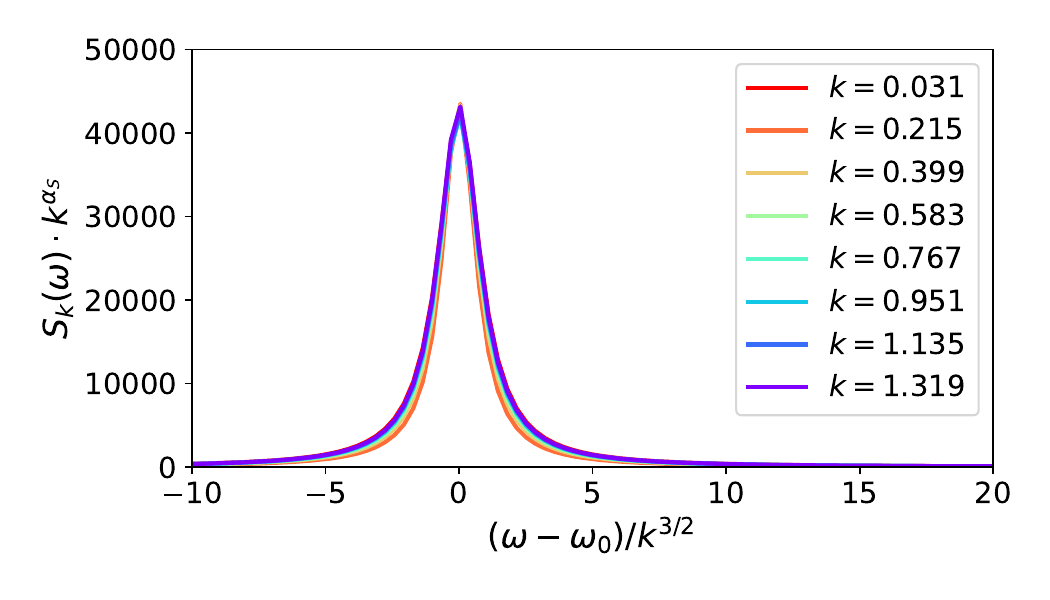}
\caption{Collapse of rescaled spectral lines $S_k(\omega)\equiv A(k,\omega-\omega(k))|_{k}$ in the variable $(\omega-\omega(k))/k^{z}$ with $z=3/2$. The spectral amplitudes are rescaled by $|k|^\alpha$ with  $\alpha\approx1.89$. Calculations are done for the  parameters in Eq.~\eqref{param} and the system length is taken as $L=2^{10}$.}
\label{fig_z_from_spectrum}
\end{figure}

\section{Quantum Dynamics}
\label{sec_quantumdyn}

\subsection{Tensor network method}
Tensor-network methods provide an efficient description of open quantum many-body systems by representing the vectorized density operator as a matrix-product state and the Liouvillian as a matrix-product operator (MPO)~\cite{ZwolakVidal2004,ProsenZnidaric2009_JSTAT}.
This approach, unlike purified representations of density matrix~\cite{verstraete04, Cuevas_2013}, does not guarantee the Hermiticity or positive-definiteness of the density matrix.
While the exact evolution generated by a Lindblad master equation is completely positive and trace preserving~\cite{lindblad_generators_1976}, finite-bond-dimension truncations of the vectorized density operator do not automatically preserve positivity and trace~\cite{BonnesCharrierLauchli2014_PRA,WernerJaschkeSilviEtAl2016,Finsterholz2020_Entropy}.
For Liouvillian time-evolution or optimization, this representation nonetheless significantly simplifies the application of the dissipative Kraus operators.
We also note another alternative method: the quantum-trajectory approach, in which individual stochastic wave functions are evolved instead of the full density matrix~\cite{Daley2014_AdvPhys,Weimer2021_RMP,BonnesLauchli2014_SuperopVsTraj,JaschkeMontangeroCarr2018_QST}. 
Tensor-network methods have been applied to a broad range of open quantum many-body systems~\cite{Guo22,Hryniuk24,KshetrimayumWeimerOrus2017_NatComm,Moca2022_PRB,Ramusat21,Westhoff25}.

We developed a package based on the ITensor library~\cite{ITensor} to compute non-equilibrium steady states (NESSs) and their dynamical correlation functions.
Our implementation supports arbitrary finite bosonic truncations, as well as fermionic and spin degrees of freedom, within a common interface.
While finishing this work, we became aware of a similar codebase released in Ref.~\cite{MisguichLibrary}.
For a unique NESS, its vectorized density operator is obtained variationally as the zero-energy ground state of the positive-semidefinite operator $\mathcal{L}^{\dagger}\mathcal{L}$,
\begin{align}
    |\rho_{\mathrm{ss}}\rangle\rangle
    &=
    \underset{\langle\langle\rho|\rho\rangle\rangle=1}
    {\operatorname*{arg\,min}}
    \langle\langle\rho|
    \mathcal{L}^{\dagger}\mathcal{L}
    |\rho\rangle\rangle
    =
    \underset{\lVert\rho\rVert_2=1}
    {\operatorname*{arg\,min}}
    \bigl\lVert
    \mathcal{L}|\rho\rangle\rangle
    \bigr\rVert_2^2
    \;.
\end{align}
This construction follows Ref.~\cite{CuiCiracBanuls2015} and enables a Hermitian DMRG-on-MPO optimization with periodic boundary conditions.
The convergence towards the steady-state is therefore controlled by gap of the squared singular values of $\mathcal{L}$.
Related tensor-network approaches include direct optimization of the non-Hermitian Liouvillian~\cite{MascarenhasFlayacSavona2015} and the methods of Ref.~\cite{Ramusat21}. 

To evaluate dynamical correlation functions, we apply the relevant operators to the NESS and propagate the resulting Liouville-space states using three complementary backends: time-evolving block decimation (TEBD)~\cite{Vidal2004}, the time-dependent variational principle (TDVP)~\cite{Haegeman2011}, and an MPO approximation to the exponential propagator~\cite{Zaletel15,VanDamme2024}. 
This provides access to momentum-resolved retarded Green's functions and NESS spectral functions.
Related tensor-network calculations of two-time correlations and momentum-resolved spectra in open systems were presented in Refs.~\cite{Kilda2019,Wolff2020}.

While the individual algorithms are established, our contribution is their integration into a unified, symmetry-resolved workflow tailored to soft-core bosons. 
In particular, we exploit the weak $U(1)$ symmetry of the Liouvillian and the associated conservation of the superparticle number in the vectorized representation, which yields a block-sparse tensor-network representation and substantially reduces the computational cost at large bosonic occupations. 
This allows us to calculate interacting spectral functions without a weak-coupling approximation and for system sizes beyond the reach of exact diagonalization. 
At fixed local dimension and bond dimension, the time-evolution cost scales linearly with system size.
The principal limitations are the local bosonic cutoff and for some parameters the high accuracy required to obtain the strongly-entangled initial NESS; in the regimes considered here, dissipation mitigates the entanglement growth.

\subsection{Spectral function in the weakly interacting quantum regime}
\begin{figure}
\hspace{-0.3cm}
	\includegraphics[width=0.25\textwidth]{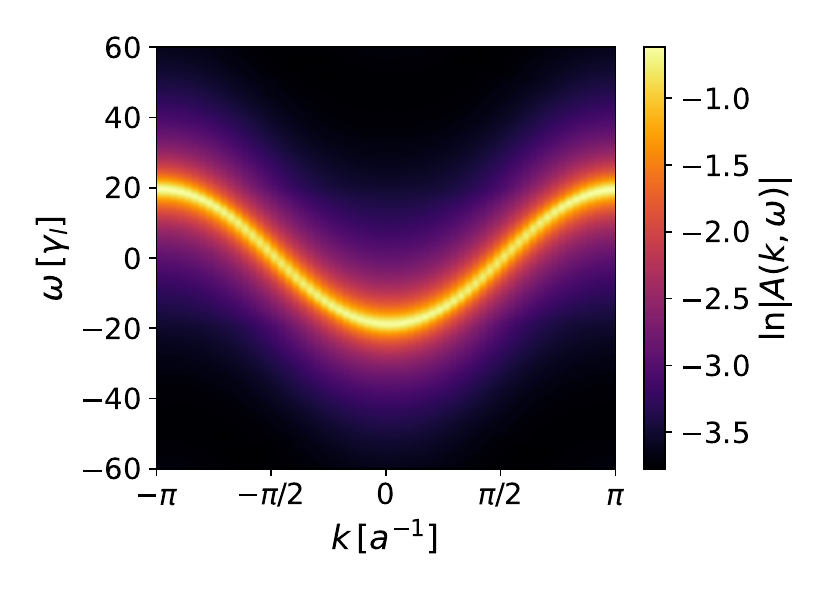}
	\hspace{-0.4cm}
\includegraphics[width=0.25\textwidth]{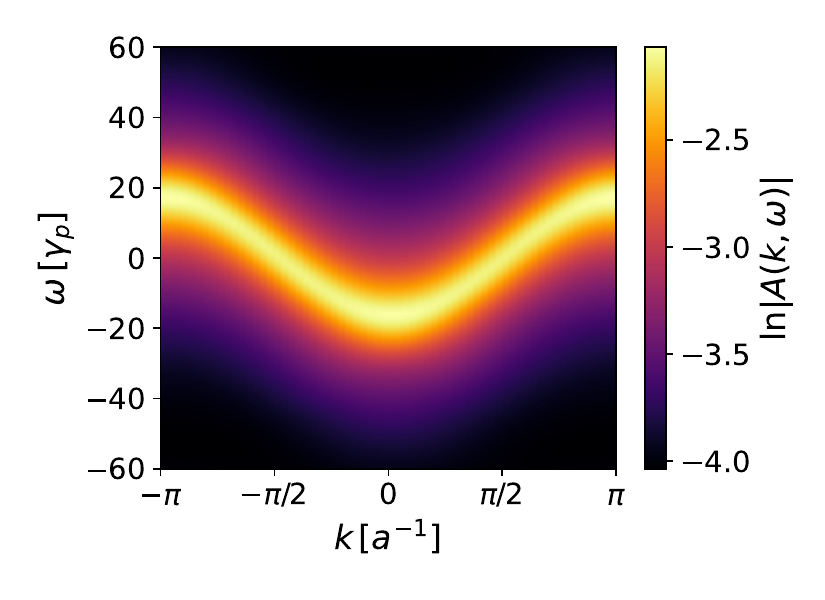}	\flushleft
	\caption{
	Tensor network prediction for the spectral function using TEBD time-evolution in the wavevector-frequency plane, with $J=10 $, $U=1$ in units indicated on the axes. Left panel: case where $\gamma_{2l}=0$, taking $\gamma_p=0.2$ at $\gamma_l=1.0$. The NESS is known exactly. Right panel: case where $\gamma_{2l}\neq 0$, taking $\gamma_l=0.5$, $\gamma_{2l}=2.0$ at $\gamma_p=1.0$. The NESS is obtained by DMRG.
	The other parameters of the simulation are $L=64$ and local Hilbert space dimension $N_s=3$ for both cases.}
	\label{fig_Aw_open}
\end{figure}

\begin{figure}
	\includegraphics[width=0.95\linewidth]{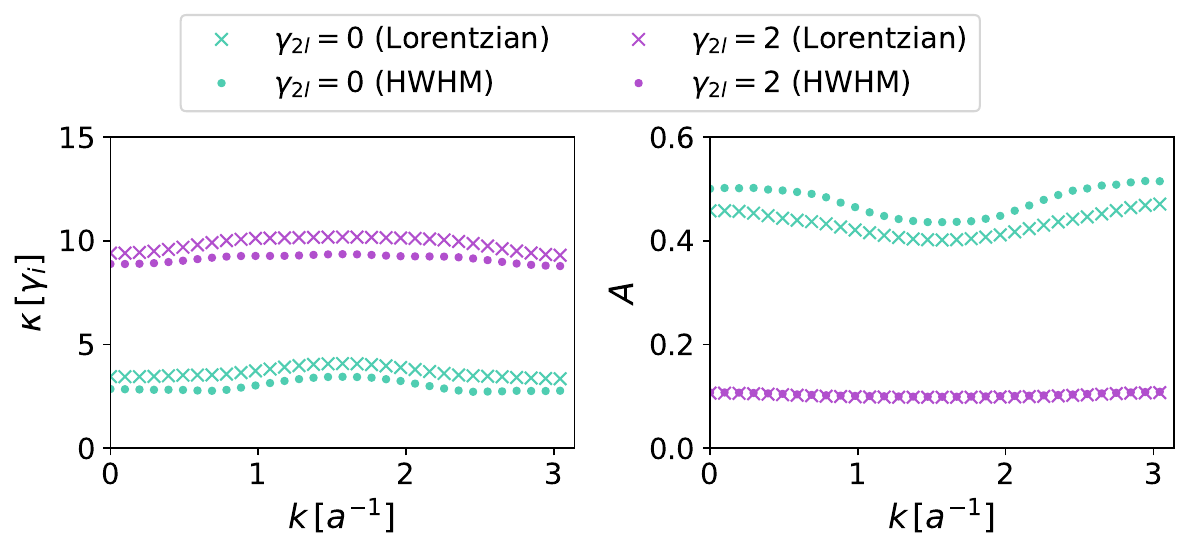}	
	\flushleft
	\caption{
    Linewidths and amplitudes of the spectral function shown Fig.~\ref{fig_Aw_open} as a function of wavevector $k$ in units of inverse lattice spacing, for both vanishing (light green colours) and non-vanishing (purple colours) two-particle loss rate $\gamma_{2l}$. 
	Two fitting methods are compared to quantify the uncertainty: a Lorentzian fit (crosses) and a half-width at half-maximum fit (HWHM, dots). The fits were performed including a constant offset for the background.}
	\label{fig_Aw_decay2}
\end{figure}

In a previous study~\cite{Zuendel25SciPost}, we calculated the spectral function $A(\omega,k)$ with exact diagonalization for a system of $L=8$ sites. The tensor network method developed in this work allows us to treat much larger system sizes with large precision in time.
We present in Fig.~\ref{fig_Aw_open} the results of the spectral function obtained by tensor network methods in the weakly interacting regime of small $U/J$ values. In this case, the observed dispersion is very close to the lattice  non-interacting one, and the main effect of tuning the loss to pump rate  is to change the linewidth and spectral weight. The linewidth and spectral amplitude depend weakly on the wavevector, both in absence and in presence of two-body losses, as shown in Fig.~\ref{fig_Aw_decay2}. 

\subsection{Linewidth from one-loop perturbative analysis}
We perform  the calculation of the linewidth  to one-loop order including two-body losses in the vacuum phase $\rho_0=0$ which leads to an analytical expression and which should be a good approximation since in the numerics $\rho$ is very low. The DeWitt notation is employed and the fields are gathered in a collective scalar field $\Phi=(\varphi_a)_{a\in\{1,\dots,4\}}=(\varphi_c,\varphi_c^*, \varphi_q, \varphi_q^*)^T$. We denote $S^{(n)}$ the $n$-th functional derivative of the action functional $S$.

The inverse propagator $G^{-1}$ is given by
\begin{align}
	G_{ab}^{-1} = G_{0,ab}^{-1} - \Sigma_{ab}\;,
\end{align}
where $G_0$ is the bare propagator at vanishing background field $G_0^{-1}=S^{(2)}[\Phi]|_{\Phi=0}$ and the self-energy $\Sigma$ to one-loop order can be derived from the background field method, and is given by
\begin{align}
	\Sigma^{\text{(1-loop)}}_{ab} = -
	\frac{\ii}{2} \frac{\delta^2}{\delta\varphi_a\delta\varphi_b} \Tr\ln \{-\ii S^{(2)}\}[\Phi]\Big|_{\Phi=0}\;.
\end{align}
We find non-zero corrections for the components
\begin{align}
	\Sigma_{14} = \Sigma_{23}^* = \frac{\gamma_0 \tilde{U}}{2\kappa_0} - \ii \tilde{\gamma}_{2l}\frac{\gamma_0}{2\kappa_0}\;,\quad
	\Sigma_{34} = -\ii\tilde{\gamma}_{2l} \frac{\gamma_0}{\kappa_0}\;,
\end{align}
such that the inverse retarded Green's function at one-loop is given by
\begin{align}
	G^{R}(\omega,k) = \Big[\omega + E(k) +\ii \kappa(k) - \Sigma_{14}\Big]^{-1}\;.
\end{align}
From the numerically fitted exponential decay of $G^R(k=0,t)$, we determine the lattice spacing $\tilde a$, that relates the couplings in the continuum to the couplings in the lattice model as $\tilde{\gamma}_{2l}=\tilde a\gamma_{2l}$, $\tilde U=\tilde a U$.

\begin{figure}[h]
    \includegraphics[width=.7\linewidth]{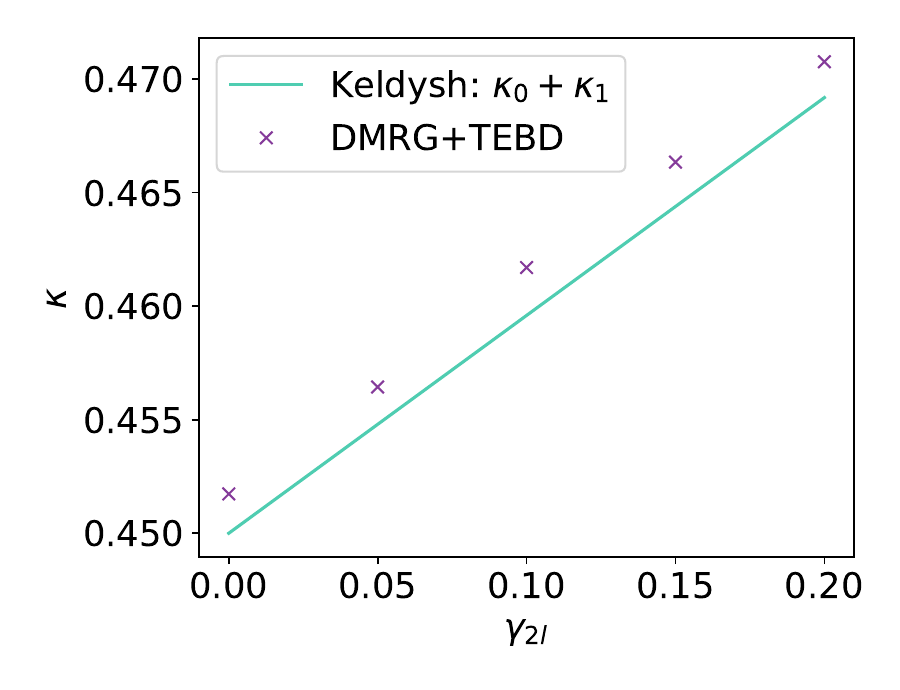}
    \flushleft
    \caption{Prediction of the temporal decay rate $\kappa$ of the retarded Greens function $G^R(k=0,t)$. The result to one-loop order in the symmetric phase is depicted in blue, the fitted decay from the numerical simulation (DMRG and TEBD) is shown in green for different values of the two-body loss $\gamma_{2l}$. Parameters of the simulation $L=3, N_s=5, J=1.0, U=0.01, \gamma_p=0.1,\gamma_l=1.0$.
    }\label{fig_decay_1loop}
\end{figure}

The imaginary part of the self-energy of the retarded Green's function $\Sigma_{14}$ is non-zero, as compared to the case, when $\gamma_{2l}=0$, at the same order in perturbation theory, and is found to be linear in $\gamma_{2l}$. As displayed in Fig.~\ref{fig_decay_1loop}, this is in perfect accordance with the  behavior of the decay rate extracted from the tensor network calculations at small $\gamma_{2l}$.

\subsection{Spectral function in the strongly interacting regime}
\begin{figure}
\hspace{-0.8cm}	\includegraphics[width=0.19\textwidth]{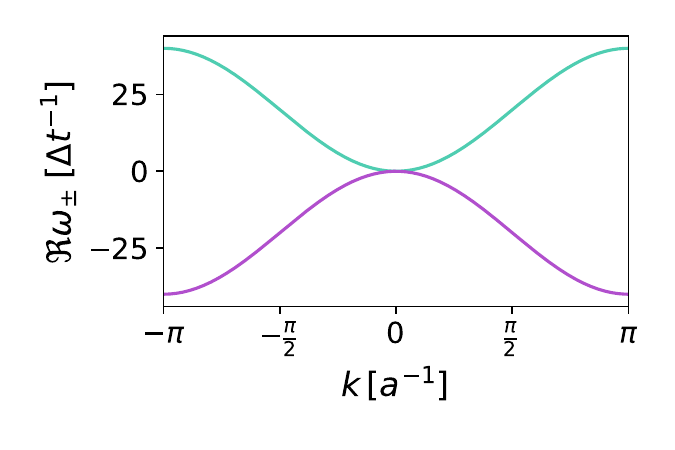}	\hspace{0.6cm}
\includegraphics[width=0.19\textwidth]{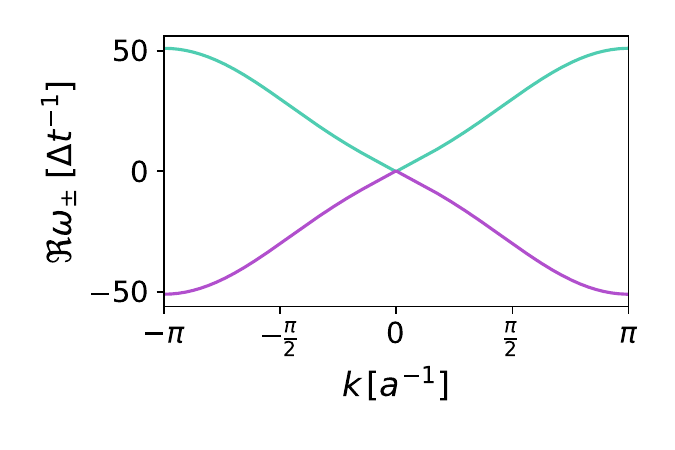} \\
	\includegraphics[width=0.235\textwidth]{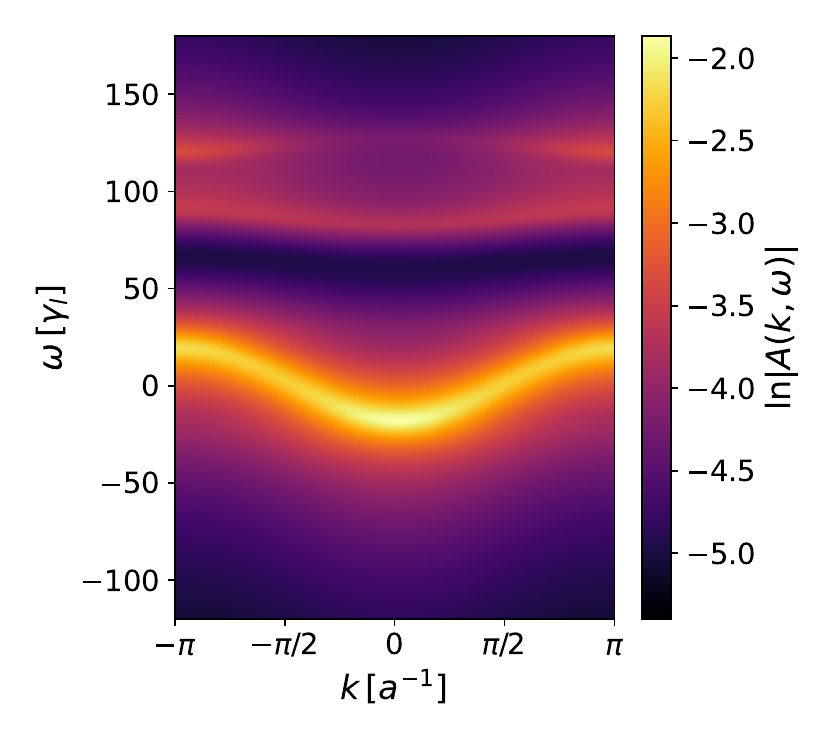} \includegraphics[width=0.235\textwidth]{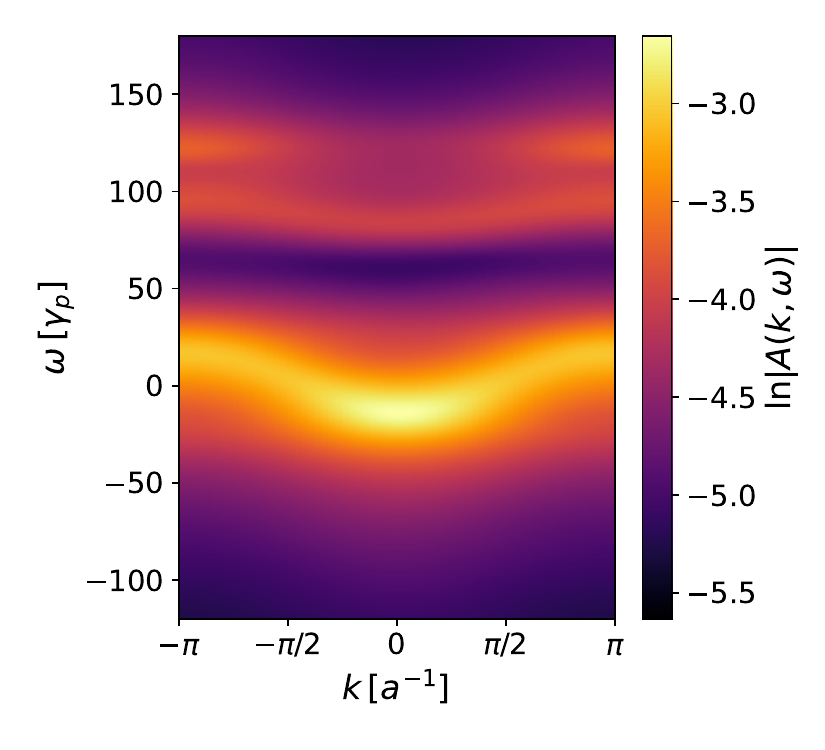}	\flushleft
	\caption{
	Mean-field dispersion (top panels) and tensor network prediction for the spectral function (bottom panels) using TEBD time-evolution in the wavevector-frequency plane, with units indicated on the axes. Left panels: case where 
	$\gamma_{2l}=0$,  taking $\gamma_p=0.2$, $\gamma_l=1.0$, $J=10 $, $U=100$. The  NESS is known exactly.
	Right panel: case where 	$\gamma_{2l}\neq 0$, taking  $\gamma_p=1.0$, $\gamma_l=0.5$,$\gamma_{2l}=2.0$ $J=10$, $U=100$. The   NESS is obtained by DMRG.
	The other parameters of the simulation are   $L=64$ and local Hilbert space dimension $N_s=3$ for both cases.}
	\label{fig_Aw_open2}
\end{figure}

We turn to the strongly interacting regime, where $U/J>1$ and $U/\gamma_i>1$, $i\in\{l,p,2l\}$. 
For simplicity, we focused on the case of local losses, i.e. we set $\gamma_d=0$.
The resulting spectral function is shown in Fig.~\ref{fig_Aw_open2}.
 
In the left panel, we display the result for an open system with only one-body loss and gain, i.e. we set $\gamma_{2l}=0$ in Eq.~\eqref{eq_lindblad}.
We work in the regime $\gamma_p<\gamma_l$, required to achieve a steady state in the infinite dimensional Hilbert space~\cite{lebreuilly_towards_2016,Zuendel25SciPost}. In this case, the mean field solution corresponds to the vacuum, $\langle \varphi_j \rangle=0$ and the elementary excitations to first order in the interactions display a cosine-like dispersion (see upper left panel). The quantum solution displays two excitation branches, a first excitation branch displaying a cosine-like dispersion, and a second excitation branch associated to doublons and occurring at energy $U$. 
 
In the right panel, we show the results in the case where additionally two-body losses are present, $\gamma_{2l}\neq 0$, in the regime  $\gamma_p>\gamma_l$. In this case, the mean-field solution predicts the emergence of a condensate, corresponding to a  $U(1)$-broken phase. The excitation spectrum on top of this mean-field state is approximately given by the Bogoliubov theory. For the parameters of the full quantum simulation, the Bogoliubov dispersion is linear for small $k$ and displays the emergence of normal (ghost) branches at positive (negative) energies, as shown in the top right panel of Fig.~\ref{fig_Aw_open2}. At strong interaction, the Bogoliubov description is expected to break down: indeed, the spectral function obtained from the tensor network calculation (bottom right panel in Fig.~\ref{fig_Aw_open2}) considerably differs from it: it does not show evidence of ghost branches, and the dispersion relation appears to be rather quadratic at small $k$. Furthermore, it displays an evident doublon branch with a larger width and flatter dispersion than in the case $\gamma_{2l}=0$. Let us emphasize that the linewidth of the lower branch display a very different character in the two cases: for $\gamma_{2l}=0$, the linewidth is essentially constant with $k$, while for $\gamma_{2l}\neq 0$, it is strongly momentum dependent, with a noticeable increase at small $k$, which is a remnant of the behavior at weak interactions predicted by the semi-classical analysis in Sec.~\ref{kpz-width}.

\begin{figure}
	\includegraphics[width=0.95\linewidth]{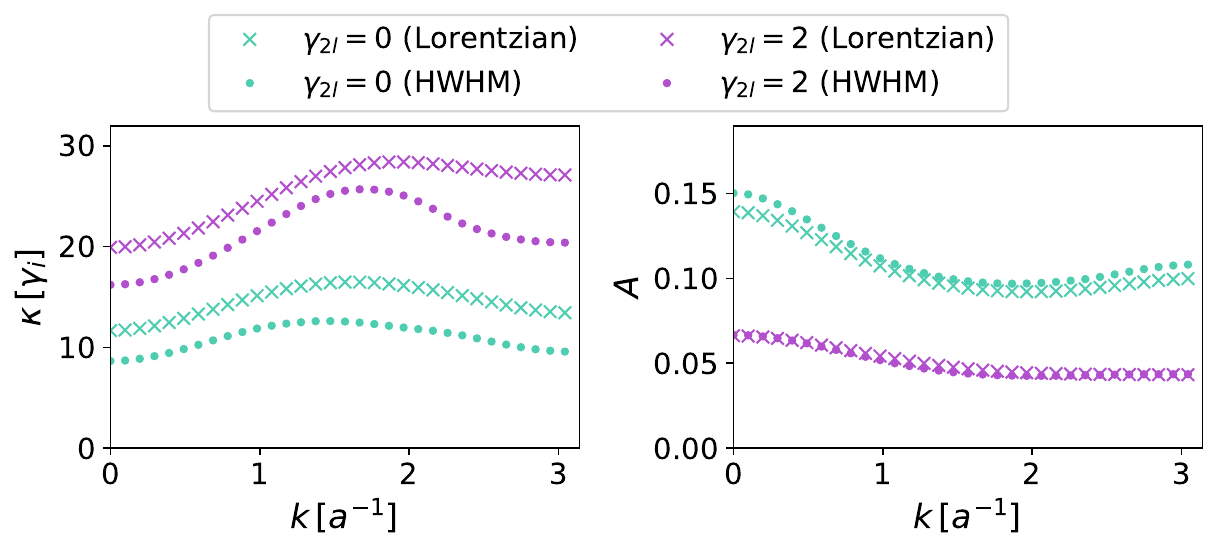}
	\flushleft
	\caption{
	Linewidths and amplitudes of the spectral function shown Fig.~\ref{fig_Aw_open2} as a function of wavevector $k$ in units of inverse lattice spacing, for both vanishing (light green colours) and non-vanishing (purple colours)  two-particle loss rate 
	$\gamma_{2l}$. The fitting methods are the same as in Fig.~\ref{fig_Aw_decay2}.}
	\label{fig_Aw_decay}
\end{figure}

\subsection{Transition in the first excitation}
In the presence of non-zero two-body losses, at increasing pump to loss ratio, the mean field description predicts the formation of a condensate above a critical pump strength. In one dimension, we do not expect a phase transition, rather a crossover behavior.
We explore here such a crossover by analyzing the spectrum of the Lindblad operator when varying the loss rate intensity from below to above the value where a transition occurs at mean field level.

\begin{figure}[h]
\centering
\begin{minipage}[t]{0.6\linewidth}
    \vspace{7pt}
    \centering
    \includegraphics[width=\linewidth]{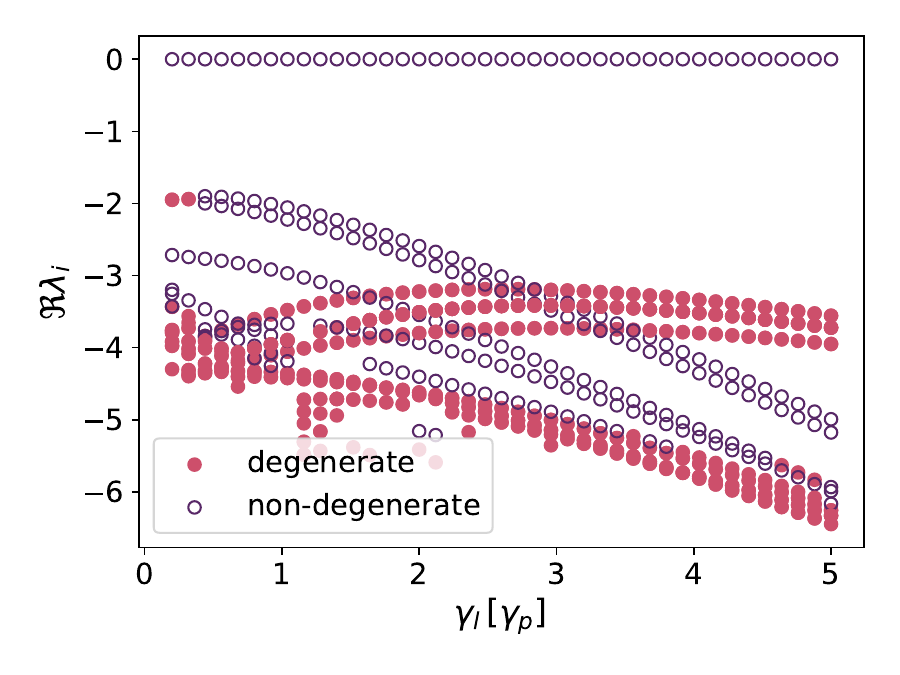}
\end{minipage}
\begin{minipage}[t]{0.38\linewidth}
    \vspace{0pt}
    \centering
    \includegraphics[width=\linewidth]{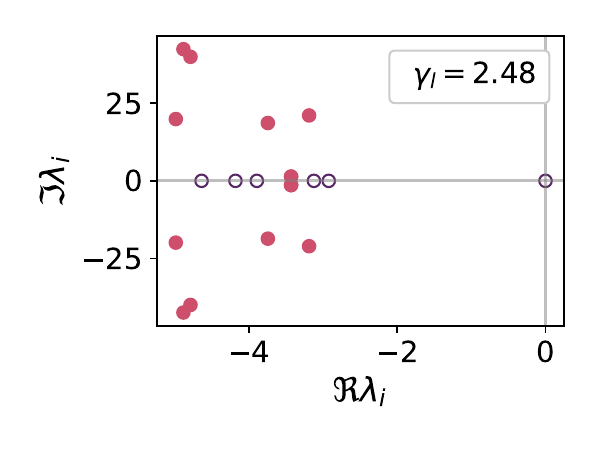}
    \includegraphics[width=\linewidth]{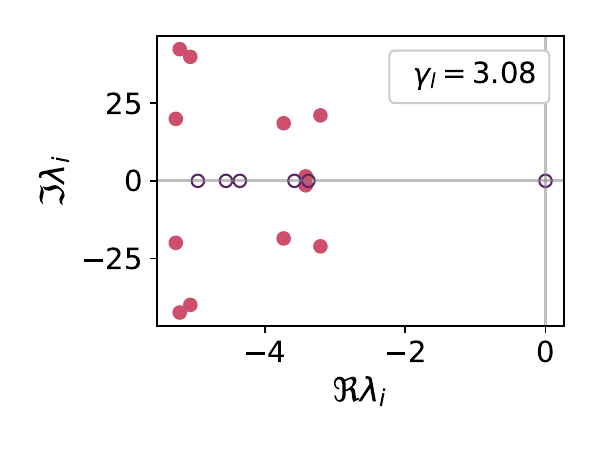}
\end{minipage}
\caption{Largest real eigenvalues from exact diagonalization  as a function of the one-body loss rate $\gamma_l$. Parameters of the simulation are $L=4, N_s=3, J=10.0, U=10.0, \gamma_p=1.,\gamma_{2l}=0.5$. Additionally, the Liouvillian spectrum for these eigenvalues is shown for two values of $\gamma_l$, showing the change of the first excitation.}
\label{fig_evs}
\end{figure}

For small system sizes, we calculate the Liouvillian spectrum  with exact diagonalization. 
In Fig.~\ref{fig_evs}, we display the lowest-energy eigenvalues. We concentrate first on the zero-energy eigenvalue, describing the steady state. 
Since the Liouvillian gap is not closing when tuning the loss rate, we confirm that there is no change in nature of the quantum steady state.
 
To characterize the structure of the two Liouvillian modes that exchange their role as the longest-lived excitation, we first determine the slowest right eigenmatrix $R_1$ from
\begin{align}
    \mathcal L R_1
    &=
    \lambda_1 R_1\;,
\end{align}
where $\lambda_1$ is the non-zero eigenvalue with the largest real part. The lifetime of this mode is therefore determined by $\operatorname{Re}\lambda_1$, independently of the overlap analysis below. Following the logic of the single-mode approximation (SMA)~\cite{Feynman1954}, generalized here to Liouville space, we construct trial excitations by applying simple observables to the non-equilibrium steady state. For an observable $O$, we construct the traceless, normalized perturbation of the non-equilibrium steady state
\begin{align}
    \chi_O
    &=
    \frac{\delta\phi_O}
    {\sqrt{\Tr\big(\delta\phi_O^\dagger\delta\phi_O\big)}}\,,
    &
    \delta\phi_O
    &=
    \frac{1}{2}
    \left\{
        O-\langle O\rangle_{\mathrm{ss}},
        \rho_{\mathrm{ss}}
    \right\}\,,
\end{align}
where $\langle O\rangle_{\mathrm{ss}}=\Tr\left(O\rho_{\mathrm{ss}}\right)$.
Its geometric similarity to $R_1$ is quantified by the normalized squared Hilbert--Schmidt overlap
\begin{align}
    \mathcal C_{\mathrm{HS}}(O)
    &=
    \frac{
        \left|
            \Tr\left(R_1^\dagger\chi_O\right)
        \right|^2
    }{
        \Tr\left(R_1^\dagger R_1\right)
    }\;.\label{eq:Liouvoverlap}
\end{align}
Since $\chi_O$ is normalized, $0\leq\mathcal C_{\mathrm{HS}}(O)\leq1$. A value close to unity means that the trial perturbation and $R_1$ have nearly the same direction in operator space. This quantity provides a geometric interpretation of the right eigenmatrix, but does not determine its lifetime or its dynamical spectral weight. For a non-normal Liouvillian, the latter is instead governed by the projection onto the corresponding left eigenmatrix.

\begin{figure}[h]
    \centering
    \includegraphics[width=.5\linewidth]
    {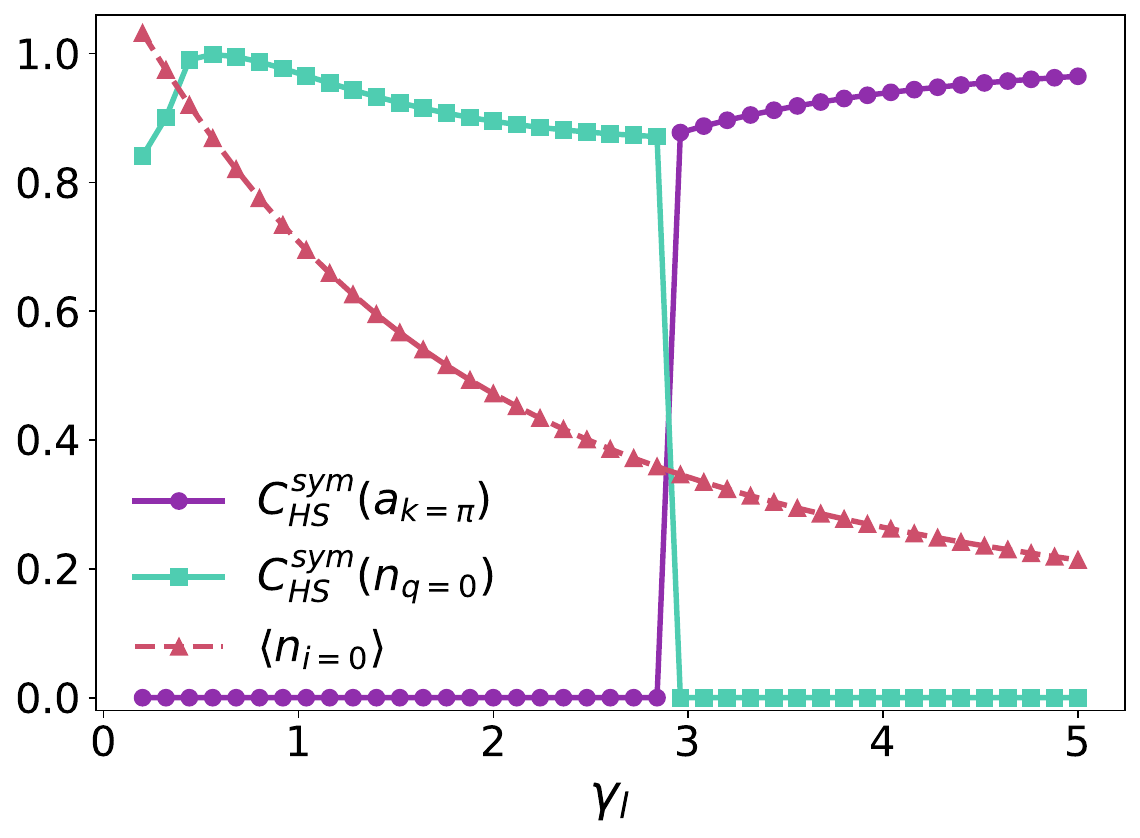}
    \caption{
        Normalized squared Hilbert--Schmidt overlap
        \(\mathcal C_{\mathrm{HS}}\) between the slowest right Liouvillian
        eigenmatrix \(R_1\) and the normalized perturbations of the NESS
        generated by \(n_{q=0}\) and \(a_{k=\pi}\). The steady-state filling
        \(\langle n\rangle_{\mathrm{ss}}\) is shown for comparison. The abrupt
        change of the overlaps reflects the interchange of the slowest mode
        at the level crossing determined from the Liouvillian eigenvalues,
        whereas the filling varies smoothly. Parameters as in Fig.~\ref{fig_evs}.
    }
    \label{fig_overlap}
\end{figure}

Fig.~\ref{fig_overlap} shows the overlap in Eq.~\eqref{eq:Liouvoverlap} for the uniform density mode
\begin{align}
    n_{q=0}
    &=
    \frac{1}{\sqrt L}
    \sum_{j=0}^{L-1}n_j
    =
    \frac{N_{\mathrm{tot}}}{\sqrt L}
\end{align}
and the staggered annihilation mode
\begin{align}
    a_{k=\pi}
    &=
    \frac{1}{\sqrt L}
    \sum_{j=0}^{L-1}
    (-1)^j a_j\;.
\end{align}
For $\gamma_l<\gamma_{l,\mathrm{cross}}$, the slowest right eigenmatrix has a near-unity overlap with the perturbation generated by $n_{q=0}$. Its operator-space structure is therefore well described by a spatially uniform filling fluctuation. For $\gamma_l>\gamma_{l,\mathrm{cross}}$, it instead closely matches the perturbation generated by $a_{k=\pi}$, and thus acquires a staggered, single-particle annihilation-like structure. The abrupt interchange of the overlaps reflects the change in the identity of $R_1$ at the level crossing previously established from the Liouvillian eigenvalues. By contrast, the steady-state filling,
\begin{align}
    \langle n\rangle_{\mathrm{ss}}
    &=
    \frac{1}{L}
    \sum_{j=0}^{L-1}
    \Tr\left(n_j\rho_{\mathrm{ss}}\right)\;,
\end{align}
decreases smoothly with increasing loss rate. The level crossing and the associated change in the structure of the slowest relaxation mode are therefore not accompanied by a sharp transition of the steady state.

Summarizing the above result, we showed that the nature of the first  excitation in the system changes at increasing loss rate from a collective, purely dissipative behavior to a coherent staggered one-particle annihilation. This change of behavior may hint at the quasi-condensate building in the weakly-interacting semi-classical limit.

\section{Conclusions and outlook}
\label{sec_conclusions}
In this paper, we combined numerical and perturbative theoretical approaches to investigate the driven-dissipative Bose-Hubbard model within different regimes of interaction strength and filling. 

In the semiclassical limit, holding at weak interactions and large filling, we identified the parameters for which the phase dynamics is governed by the KPZ fixed point, both in space-time correlations as well as in the statistics of the fluctuations in the phase increments. We determine for these parameters a minimum size and filling of the system  for the KPZ scaling to become observable.
In such a regime, we have shown that KPZ scaling is visible in the spectral function through the analysis of the linewidth of the excitation dispersion: at difference from the predictions of the Bogoliubov theory, the spectral lines collapse onto a single one according to the KPZ scaling and the linewidth grows as $\sim k^z$ with $z=3/2$ the dynamical critical exponent of the KPZ universality class. 

In the deep quantum regime, corresponding to small filling and large interactions, we developed a tensor network method capable of computing the steady state and the real-time dynamics of the one-dimensional driven-dissipative bosons. This method is very accurate at low filling and for low entanglement values characteristic of the open lossy case under consideration, and scales linearly in computation time with the system size.
We have compared the spectral function at weak and strong interactions, subjected simply to one-body loss and pump or an additional two-body loss. While at weak interactions the linewidth is weakly momentum dependent, it varies strongly for larger interaction strength.
Furthermore, using the Keldysh formalism, we have performed a perturbative calculation to one-loop order to obtain the width of the spectral lines, finding an excellent agreement with our numerical results at low interactions.

Next, we have established that the Liouvillian gap is not closing where the semi-classical transition takes place in higher dimensions, but that the longest living excitations in the Liouvillian spectrum are changing in nature by tuning the one-body loss rate compared to the pump rate. We have identified the nature of excitations by calculating the matrix elements between the NESS and the excited state and highlighted that at increasing pump the nature of the excitations becomes more collective, hinting to the crossover predicted in the semiclassical limit.

In this study, we presented the analysis of two complementary parameter regimes: on the one hand, the regime characterized by large filling and weak interactions, where the semi-classical method provides a good description, and on the other hand the weak filling and arbitrary interactions, where quantum effects play a major role and the semi-classical picture breaks down. The study of larger fillings at intermediate and large interactions remains an open challenge both for theory and simulation. 

\begin{acknowledgments}
This work is part of HQI (www.hqi.fr) initiative and is supported by France 2030 under the French National Research Agency grant number ANR-22-PNCQ-0002.
L.H. acknowledges the Tremplin funding from CNRS Physique and was also supported by the ANR JCJC ANR-25-CE30-2205-01. 
\end{acknowledgments}

\appendix

\section{Density distributions in classical and quantum regimes}\label{sec_qm_to_cl}
In this appendix, we directly compare the density distributions obtained within the semiclassical and fully quantum descriptions. This provides a connection between the two regimes studied in this work and allows us to examine how the same dissipative processes manifest themselves when quantum fluctuations are fully retained. We use the parameters given in Eq.~\eqref{param}, apart from the varying pump strength.

\begin{figure}[h]
\centering
\hspace{-0.9em}
\includegraphics[width=1.02\linewidth]{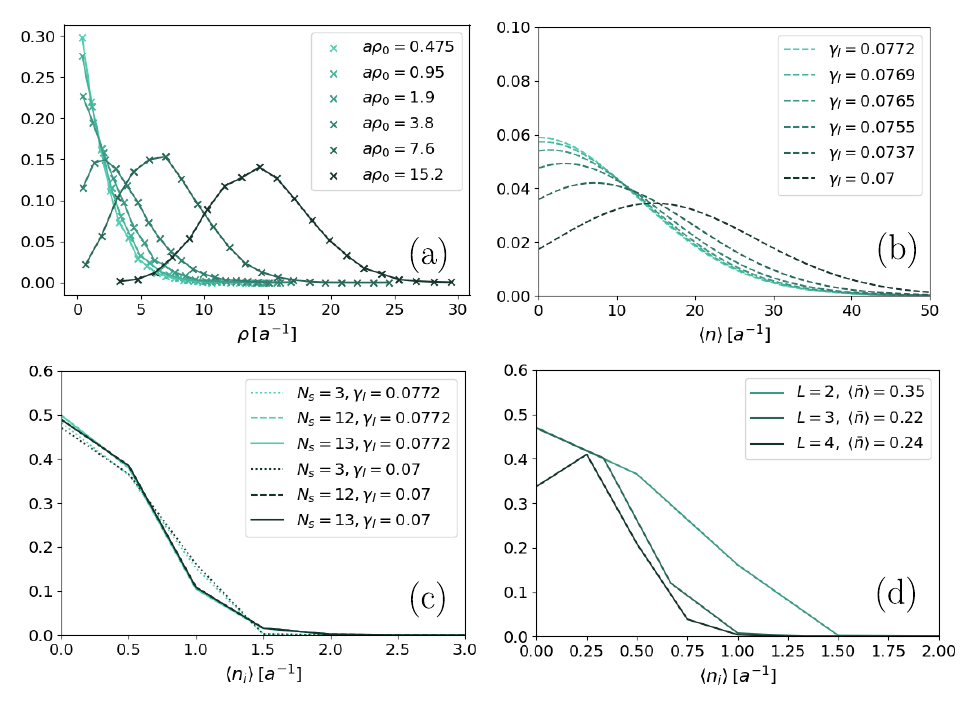}
\caption{Comparison of the predictions of semiclassical and quantum calculations. (a) Density distribution in semi-classical evolution  and (b) exact weights of the steady state density matrix of zero-dimensional ($L=1$) quantum system. Parameters are given in Eq.~\eqref{param}, local Hilbert space truncation $N_s=100$ in Fig.~(b). In both figures the loss rate $\gamma_l$ is tuned as indicated in panel (b). In panel~(a), we indicate in the labels the mean field prediction of the particle density.
Subfigures~(c) and~(d) show the weight of the different particle number sectors of the steady state density matrix rescaled by the system size. In panel~(c) for $L=2$ at two different loss rates and for several different Hilbert space truncations $N_s$, whereas in (d), we display the change of the distribution with the system size $L$ at fixed $\gamma_l=0.07$ and $N_s=3$. Overall, we observe that for extended systems there is a substantially lower occupation in the quantum simulation than in the semi-classical counterpart.}
\label{fig_desnity_stat}
\end{figure}

Of particular importance is the non-local loss with rate $\gamma_d$. In the semiclassical description, this term gives rise to a dissipative contribution to the diffusion coefficient of the driven-dissipative Gross--Pitaevskii equation and is crucial for numerically accessing the KPZ regime. We therefore investigate how the same non-local loss affects the steady state of the fully quantum lattice model. To this end, we compare the complete density distributions obtained from the semiclassical stochastic evolution and from exact diagonalization of the quantum Liouvillian. The results are shown in Fig.~\ref{fig_desnity_stat}.

Panel~(a) displays the semiclassical density distributions, with the statistics collected over different noise realizations at a fixed space-time point. As the condensate density decreases, the distribution changes from a Gaussian form, as expected for coherent states, to an approximately exponential form.
Upon approaching from below the mean-field threshold $\gamma_p=\gamma_l$, the average density decreases continuously and reaches values close to zero. Note that the pump strength is tuned in panels (a) and (b) in the same way, indicated in panel~(b).

We show the results of the full quantum model in panels~(b)--(d). We display the occupation probability resolved by number sector as extracted from the quantum steady-state density matrix.
Specifically, we calculate the weight $p_N=\operatorname{Tr}(P_N\rho_{\mathrm{ss}})$ associated with each particle-number sector $N$ and plot it as a function of the corresponding density $N/L$. Panel~(b) shows the results for a single lattice site, $L=1$, at the same one-body loss rates as in panel~(a). Increasing $\gamma_l$ shifts the distribution towards lower occupations, but does not substantially change its shape. In particular, the exact quantum NESS retains a finite occupation close to the mean-field threshold.

Crucially, the single-site problem does not capture the effect of the non-local loss. The corresponding jump operator is proportional to the difference between the fields on neighboring sites and therefore vanishes identically for $L=1$. Consequently, the results in panel~(b) are independent of $\gamma_d$.

The situation changes qualitatively as soon as the system is extended. For $L\geq2$, the non-local jump operators become nonzero and damp finite-momentum modes, while the spatially uniform mode remains unaffected. Quantum fluctuations populate these nonuniform modes, so that the non-local dissipation provides an additional channel through which particles are removed from the system. This explains the pronounced shift towards lower densities between the single-site result in panel~(b) and the $L=2$ result in panel~(c).

We can observe at $L=2$, that changing the one-body loss rate, and hence the distance from the mean-field threshold, has only a weak effect on the distribution. We therefore show only the smallest and largest values of $\gamma_l$ in panel~(c). This panel also compares different local Hilbert-space truncations. The agreement between $N_s=12$ and $N_s=13$ demonstrates convergence, while already $N_s=3$ reproduces the qualitative form of the distribution, which we exploit  in the next panel in order to study slightly larger system sizes: panel~(d) shows the evolution of the distribution with system size at fixed $\gamma_l=0.07$ and $N_s=3$, with the corresponding mean occupations $\langle\bar n\rangle$ indicated in the panel.

This direct comparison reveals that the same microscopic parameters lead to substantially different density regimes in the two descriptions. In the extended quantum system, the mean occupation is reduced to approximately $\langle\bar n\rangle\simeq0.2$ and depends only weakly on the distance from the mean-field threshold. By contrast, the semiclassical density changes strongly upon approaching the threshold and reaches the large occupations for which KPZ scaling can be observed.

\bibliographystyle{apsrev4-2}

\bibliography{arxiv}

\end{document}